\documentclass[prd,preprint,nofootinbib]{revtex4}

\usepackage{graphicx}
\usepackage{amssymb}
\usepackage{amsmath}
\usepackage{color}

\begin{document}
 
\newcommand{\be}{\begin{equation}}
\newcommand{\ee}{\end{equation}}
\newcommand{\aprime}{\mathbf{a}^{\prime}}
\newcommand{\bprime}{\mathbf{b}^{\prime}}
\newcommand{\kh}{\hat{k}}
\newcommand{\Ip}{\vec{I}_+}
\newcommand{\Imi}{\vec{I}_-}
\newcommand{\bc}{\begin{cases}}
\newcommand{\ec}{\end{cases}}
\newcommand{\cD}{\mathcal{D}}
\newcommand{\xv}{\mathbf{x}}
\newcommand{\qv}{\mathbf{q}}
\newcommand{\pv}{\mathbf{p}}
\newcommand{\trec}{t_{\mathrm{rec}}}
\newcommand{\trei}{t_{\mathrm{rei}}}

\newcommand{\red}{\color{red}}
\newcommand{\cyan}{\color{cyan}}
\newcommand{\blue}{\color{blue}}
\newcommand{\magenta}{\color{magenta}}
\newcommand{\yellow}{\color{yellow}}
\newcommand{\green}{\color{green}}
\newcommand{\rem}[1]{{\bf\blue #1}}

\begin{flushright}
ICRR-Report-567-2009-29 \\
IPMU 10-0046 \\
\end{flushright}


\title{
B-mode polarization induced by gravitational waves from\\ kinks on infinite cosmic strings
}


\author{Masahiro Kawasaki$^{(a, b)}$, Koichi Miyamoto$^{(a)}$ 
and Kazunori Nakayama$^{(c)}$}

\affiliation{%
$^a$Institute for Cosmic Ray Research,
     University of Tokyo, Kashiwa, Chiba 277-8582, Japan\\
$^b$Institute for the Physics and Mathematics of the Universe, 
     University of Tokyo, Kashiwa, Chiba 277-8568, Japan\\
$^c$KEK Theory Center, Institute of Particle and Nuclear Studies, KEK, 
     Tsukuba, Ibaraki 305-0801, Japan
}

\date{\today}

\vskip 1.0cm

\begin{abstract}
 We investigate the effect of the stochastic gravitational wave (GW) background produced by kinks on infinite cosmic strings, whose spectrum was derived in our previous work, on the B-mode power spectrum of the cosmic microwave background (CMB) anisotropy.
 We find that the B-mode polarization due to kinks is 
 comparable to that induced by the motion of the string network
and hence the contribution of GWs from kinks is important 
for estimating the B-mode power spectrum originating from cosmic strings.
If the tension of cosmic strings $\mu$ is large enough i.e., $G\mu \gtrsim 10^{-8}$, 
B-mode polarization induced by cosmic strings can be detected by future CMB experiments.
\end{abstract}

 \maketitle

 \section{Introduction}
 
Cosmic (super)strings can be produced in the early Universe at the phase transition associated with spontaneous symmetry breaking~\cite{Vilenkin}, 
the end of supersymmetric hybrid inflation~\cite{Jeannerot:1997is, Lyth:1997pf}, or 
the end of the brane inflation~\cite{Sarangi:2002yt, Dvali:2003zj}.
They can be a clue to particle physics beyond the standard model and the history of the early Universe, which is difficult to obtain in terrestrial experiments. 
How to find signatures of cosmic strings in cosmic microwave background (CMB) experiments has been extensively discussed for decades. 
Especially, B-mode polarization of the CMB induced by the cosmic string network was investigated in many papers~\cite{Seljak:1997ii,Benabed:1999wn,Pogosian:2003mz,Bevis:2007qz,Pogosian:2007gi}.  

B-mode polarization, which has not been detected yet, is polarization of the parity-odd type. 
It cannot be produced by the primordial scalar perturbation from the inflationary era, which is widely believed to be the main origin of the present structure of the Universe.
On the other hand, the tensor perturbation can be a source of B-mode polarization. 
Some inflation models can produce the intense tensor perturbation enough to generate detectable B-mode, while others cannot.

Cosmic strings can also induce B-mode.
Cosmic strings move in the Universe in a very complicated and nonlinear way, constantly generating all types of perturbations, scalar, vector and tensor ones.
Therefore, dynamics of the cosmic string network induces B-mode and it reaches an observable level if the tension of cosmic strings, $\mu$, is large enough, say, $G\mu \gtrsim 10^{-7}$~\cite{Pogosian:2007gi}. Here, $G$ denotes the Newton constant.

In this paper, we point out that there is an additional source of B-mode when the cosmic string network exists. 
It is the stochastic gravitational wave (GW) from kinks on infinite strings.\footnote{
The effects of the kinks on infinite strings are partially reflected in the calculation in Ref.~\cite{Bevis:2007qz} based on the lattice simulation. 
However, such a simulation covers only the limited period of the evolution of the string network, so the correct kink distribution on infinite strings cannot be taken into account by this method.
Moreover, it is impossible to completely separate the GWs from kinks from those emitted at the phase transition.
Therefore, our calculation based on the kink distribution derived analytically is complementary to the calculation in Ref.~\cite{Bevis:2007qz}.
}
In the previous paper~\cite{Kawasaki:2010yi}, we investigated GWs emitted from kinks on infinitely long strings, and found that GWs with a wavelength comparable to the Hubble horizon scale are generated. 
These long wavelength GWs can produce an observable B-mode in the CMB.
As a result, we find that the contribution of such a GW background to a B-mode is comparable to that due to dynamics of the cosmic string network and detectable by future CMB experiments, such as PLANCK or CMBpol. 

This paper is organized as follows. 
In section 2, we present the formalism which we adopt in order to compute the BB power spectrum. 
In section 3, we briefly review the result of Ref.~\cite{Kawasaki:2010yi} 
for the GW spectrum from kinks which will be used in the following analysis.
In section 4, we show the resulting BB power spectrum and discuss its observational implications. 
Section 5 is devoted to summary.

  \section{Formalism for Computation of the BB Power Spectrum}

In this paper, we adopt the formalism described in Ref.~\cite{Weinberg}.
The tensor mode of the metric perturbation $D_{ij}$ is defined as $g_{ij}(t,\xv)=a(t)^2(\delta_{ij}+D_{ij}(t,\xv))$, where $a(t)$ is the scale factor, and it is symmetric, transverse and traceless: $D_{ij}=D_{ji}, \  \partial_iD_{ij}=0, \ D_{ii}=0$.
We expand $D_{ij}$ in the following form,
\be
	D_{ij}(t,\xv)\equiv\int d^3q e^{i\qv\cdot\xv}D_{ij}(t,\qv)\equiv\sum_{\lambda=\pm 2}  \int d^3q e^{i \qv\cdot\xv}e_{ij}(\hat{q},\lambda)\cD(t,\qv,\lambda),
\ee
where $\hat{q}\equiv \qv/|\qv|$, $\lambda$ denotes the helicity of the GW and $e_{ij}(\hat{q},\lambda)$ is the polarization tensor. 
We can think of $\cD(t,\qv,\lambda)$ as a stochastic variable since it is the product of random GW emission by kinks on infinite strings.
Its root mean square is inferred from previous paper~\cite{Kawasaki:2010yi} and given in the next section.

As described in~\cite{Weinberg}, this metric perturbation relates to the polarization of photons.
We do not explain the detail here, but write down only several important equations.
The BB power spectrum, $C_{BB,\ell}$, is defined by $\langle a_{B,\ell m}^* a_{B,\ell ' m'} \rangle = C_{BB,\ell} \delta_{\ell \ell '}\delta_{mm'}$ ~\cite{Seljak:1996gy},
where $a_{B,\ell m} $ is some combination of the coefficients of the multipole expansion of the Stokes parameters, $Q(\hat n)$ and $U(\hat n)$, by the spin-weighted harmonics. 
$a_{B,\ell m}$ can be written as
\be
	a_{B,\ell m}=i^{\ell}T_0\sqrt{\frac{\pi(2\ell+1)}{8}}\sum_{\lambda=\pm 2}\pm\int d^3q D^{(\ell)}_{m,\lambda}(S(\hat{q})) \int ^{t_0}_0 dt P(t) \Psi(t,\qv,\lambda) \left[ \left( 8\rho + \rho^2 \frac{\partial}{\partial \rho} \right)\frac{j_{\ell}(\rho)}{\rho^2}\right] \bigg|_{\rho=qr(t)} , \label{aBlm}
	\ee
where $T_0$ is the present CMB temperature, $t_0$ is the age of the Universe,  $r(t)=\int^{t_0}_{t}\frac{dt^{\prime}}{a(t^{\prime})}$, $D^{(\ell)}$ is the spin-$\ell$ unitary representation of the rotation group, $S(\hat{q})$ is the rotation which takes the three-axis into the direction $\hat{q}$ and $j_{\ell}$ is the $\ell$-th spherical Bessel function.
$P(t)=\omega_c(t) \exp(-\int^{t_0}_{t}\omega_c(t^{\prime})dt^{\prime})$ is the so-called visibility function, which has sharp peaks at the moment of recombination and reionization.
$\omega_c(t)$ is the rate of Thomson scattering.
$\Psi$ is the function, which satisfies
\be
	 \Psi  (t,\qv,\lambda)= 
	 \frac{3}{2}\int^t_0 dt^{\prime} e^{-\int^t_{t^{\prime}}dt^{\prime\prime}\omega_c(t^{\prime\prime})}
	\left[ -2\dot{\cD}(t^{\prime},\qv,\lambda)
	K\left(q\int^t_{t^{\prime}}\frac{dt^{\prime\prime}}{a(t^{\prime\prime})}\right)
	+\omega_c(t^{\prime})F\left(q\int^t_{t^{\prime}}
	\frac{dt^{\prime\prime}}{a(t^{\prime\prime})}\right)
	\Psi(t^{\prime},\qv,\lambda)\right], \label{Psieq}
\ee
\be
	K(x)=j_2(x)/x^2,\ F(x)=j_0(x)-2j_1(x)/x+2j_2(x)/x^2,
\ee
The definitions of the Stokes parameters, $a_{B,\ell m}$ and $\Psi$ are found in~\cite{Weinberg}.
If we know the way for $\cD$ to evolve precisely, we can get $\Psi$ through Eq.~(\ref{Psieq}) and calculate the power spectrum by integrating Eq.~(\ref{aBlm}).
However, we cannot know the phase of $\cD$, which varies randomly, since the stochastic background of GW is formed by random and continuous accumulation of GWs from kinks.
We can find only the expectation value of its amplitude.
Nevertheless, we can estimate the BB power spectrum using the $\delta$-function-like property of 
$P(t)$, as described in the Appendix.    
The B-mode power spectrum is calculated as
\be
	C_{BB,\ell}\simeq
	\pi^2T_0^2\int dqq^2\left[ A_\ell(q)\dot{\tilde{\cD}}^2(t_{\mathrm{rec}},q)
	+B_\ell(q)\dot{\tilde{\cD}}^2(t_{\mathrm{rei}},q)\right], \label{CBB}
\ee
where $\dot{\tilde{\cD}}$ is defined in the next section, $A_\ell(q)$ and $B_\ell(q)$ are defined in Appendix, and $t_{\mathrm{rec}}(t_{\mathrm{rei}})$ is the 
cosmic time at recombination(reionization), where the visibility function has a peak.  
 
  \section{Spectrum of the Stochastic Gravitational Wave Background}

The amplitude of the tensor perturbation is found as below. 
In~\cite{Kawasaki:2010yi}, we derived the spectrum of the stochastic GW background,
using this kink distribution function~\cite{Copeland:2009dk}, which describes the abundance of kinks for a given sharpness. 
In the matter-dominated(MD) era, the energy density of GWs of frequency $\sim \omega$ is
 \begin{align}
	 \frac{d\rho}{d\ln\omega}  \sim 
 	\bc
	10G\mu^2(\omega t)^{C_m}t^{-2} 
	&{\rm for~~}t^{-1}<\omega<\omega^{\rm (MD)}_1(t) = \left( \frac{t_{\rm eq}}{t}\right)^{A_m}t^{-1} \\
	\displaystyle
	10G\mu^2 \left(\frac{t_{\rm eq}}{t} \right)^{-2D/A_r}(\omega t)^{C_r} t^{-2}
	&{\rm for~~} \omega^{\rm (MD)}_1(t)< \omega < \omega^{\rm (MD)}_2(t)=\left(\frac{t_{\rm eq}}{t} \right)^{A_m} \left(\frac{t_r}{t_{\rm eq}} \right)^{A_r}t^{-1}
    \ec, \label{powtotmat2}
\end{align}
where $A_m=-0.8$, $A_r=-0.92$, $C_m=-0.17$, $C_r=0.14$, $D=0.11$, 
$t_{\rm eq}$ is the 
cosmic time of the matter-radiation equality and $t_r$ is the time when the reheating completes.
GWs with frequency larger than $\omega^{\rm (MD)}_2 (t)$ are irrelevant because their wavelength is too short to affect the large-scale density perturbation probed by CMB observations.   
The periods concerning the CMB polarization are only those around the recombination and the reionization, hence it is sufficient to consider the matter-dominated era only.
This GW background consists of GWs emitted toward random directions from random points in the Universe at random time.
Therefore, we can think of it as being isotropic and homogeneous.\footnote{
	In Ref.~\cite{Kawasaki:2010yi}, we omitted GWs which do not overlap others from the 	
	``background'', following the prescription given Ref.~\cite{Damour:2001bk}.
	Here, however, we do not consider this subtlety and include all GWs in the background, 
	because we are paying attention to only long wavelength modes, most of which overlap others.
}

We want to connect this expression of the energy density to the amplitude of GWs $\cD$.
At the small scale where the cosmic expansion can be neglected, the energy density of GWs can be written as~\cite{Maggiore:1999vm}
\be
	\rho=\frac{1}{32\pi G}\langle \dot{D}_{ij}(t,\mathbf{x})\dot{D}^{ij}(t,\mathbf{x})\rangle. 
	\label{densityGW}
\ee
This expression applies to GWs from infinite strings, whose wavelength is shorter than the Hubble radius.
Under the present notation, the energy density of GWs whose frequency $\sim\omega$ can be written as
\be
	\frac{d\rho}{d\ln\omega}=
	\frac{1}{2G}\omega^3a^3\dot{\tilde{\cD}}^2(t,q). \label{drhodomega}
\ee
Here, we set $\langle \dot{\cD}(t,\qv,\lambda)\dot{\cD}(t,\qv^{\prime},\lambda^{\prime})\rangle\equiv\dot{\tilde{\cD}}^2(t,q)\delta_{\lambda\lambda^{\prime}}\delta^3(\qv-\qv^{\prime})$ and used the fact that $\cD(t,q)$ is oscillating with frequency $q/a=\omega$.
Eventually, from Eqs.~(\ref{powtotmat2}) and (\ref{drhodomega}) we obtain 
 \be
	 \dot{\tilde{\cD}}(t,q)  \sim 
	 \begin{cases}
	\sqrt{20}G\mu q^{-3/2}(\omega t)^{C_m/2} t^{-1}& 
	\mathrm{for~~} t^{-1}<\omega< \omega^{\rm (MD)}_1(t) \\
	\displaystyle
 	\sqrt{20}G\mu q^{-3/2} \left(\frac{t_{\rm eq}}{t} \right)^{-D/A_r}(\omega t)^{C_r/2}t^{-1}  
	& \rm{for~~} \omega^{\rm (MD)}_1(t)< \omega < \omega^{\rm (MD)}_2(t) 
	\end{cases} .
	\label{Dtq}
\ee

  \section{BB power spectrum}

Now let us calculate the BB power spectrum.
As for the cosmological parameters,
we used the result of the 7-year WMAP observation~\cite{Jarosik:2010iu}.
Besides, we have to specify the ionization history, or the shape of $\omega_c(t)$.
It is given as $\omega_c(t)=\sigma_{\rm T} n_e(t)$, where 
$\sigma_{\rm T}$ is the Thomson scattering cross section 
and $n_e$ is the number density of electrons.
We calculate the time evolution of $n_e(t)$ around the recombination epoch
by using the RECFAST code~\cite{Seager:1999bc,recfast}.
Concerning the reionization, we make an approximation that the reionization occurs suddenly at some redshift $z_{\rm re}$ : $n_e$ jumps from $0$ to some value $n_{e0}$ at $z_{\rm re}$.
Thereafter, $n_e$ decreases in proportion to $(1+z)^{-3}$.
In short, we assume
\be
	\omega_c(t)=
	\bc
	0 & \mathrm{for~~}  z>z_{\rm re} \\
	\omega_{c0}(1+z)^{-3} & \mathrm{for~~} z<z_{\rm re} \\
	\ec ,
\ee
around the reionization.
We set $z_{\rm re}=10.4$ according to Ref.~\cite{Jarosik:2010iu}.
$\omega_{c0}$ is a constant which is determined so that $\int^{t_0}_{t_{\rm rei}} dt \omega_c(t)$ 
conforms to the reionization optical depth $\tau=0.087$~\cite{Jarosik:2010iu}.        

\begin{figure}[htbp]
\begin{center}
\includegraphics[width=100mm]{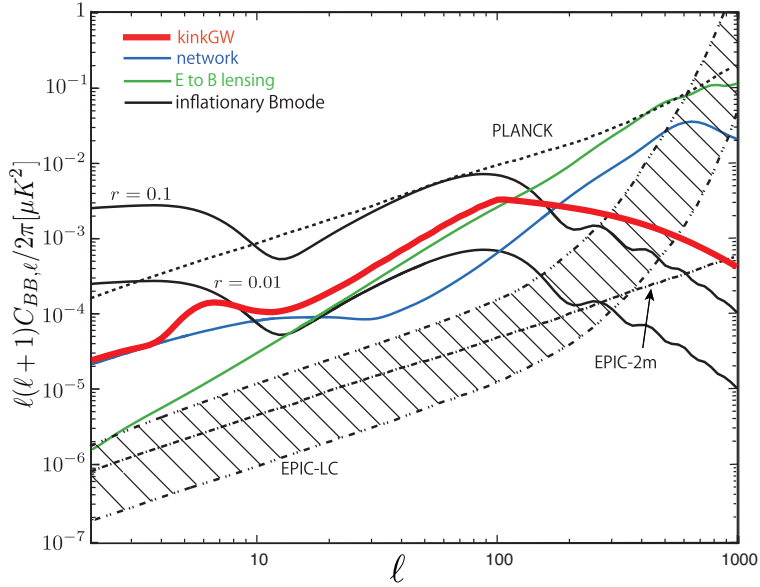}
\caption{The BB power spectrum induced by various processes and the sensitivity curves of the future CMB experiments. 
The sensitivity curves are derived from Ref.~\cite{Baumann:2008aq}. 
 }
\label{fig:CBB}
\end{center}
\end{figure}

We show the resulting BB power spectrum in Fig.~\ref{fig:CBB} (thick red) 
with the spectra produced by the string network dynamics (blue),  
the inflationary tensor perturbation with tensor-to-scalar ratio of 0.1 and 0.01 (black)
and the lensing effect (green). 
We also show the sensitivity curves of PLANCK and  two different realizations of planned CMBpol satellites, EPIC-LC and EPIC-2m.
The spectrum induced by the string network dynamics is drawn by CMBACT
\cite{Pogosian:1999np,CMBACT} 
and those originating from the inflation are obtained by the CAMB code~\cite{Lewis:1999bs}.
The value of $G\mu$ is set to be $10^{-7}$, which is close to the present 
observational upper bound~\cite{Wyman:2005tu,Seljak:2006bg,Bevis:2007gh}.
In computation using CMBACT, the network parameters are set as follows : 
the wiggliness $\alpha_r=1.8$, the r.m.s. string velocity $v_r=0.64$, and the ratio of the correlation length to the cosmic time $\gamma_r=0.3$, where subscript $r$ means the values in the radiation era.
They are derived from the results of the simulations~\cite{Bennett:1989yp,Allen:1990tv,Martins:2003vd,Martins:2005es}, and extrapolated to the matter era by the procedure in the code of CMBACT.
 
We can see the spectrum induced by the GWs from kinks on infinite strings has two peaks, one of which is located at $\ell\sim100$ and the other at $\ell\sim5$.
The peak at $\ell\sim 100$($\ell\sim 5$) is induced by GWs which exist at the recombination(reionization).
At every moment, the lower limit of frequency of existing GWs is roughly the Hubble parameter at that time.
Besides, the amplitude of GWs declines toward higher frequency.
As a result, the position of each peak is set by the Hubble parameter at the recombination or the reionization. 
Remembering that GWs of frequency comparable to the Hubble parameter are, as discussed in \cite{Kawasaki:2010yi}, emitted by new kinks, one finds that the peak at $\ell\sim 100(\ell\sim 5)$ is due to GWs emitted slightly before the recombination(reionization) by kinks which are born a little before the recombination(reionization).
Since the BB spectrum by kinks is comparable to or somewhat grater than that by network dynamics in some regions, the total BB power spectrum induced by cosmic strings is deformed by the effect of the GWs from kinks on infinite strings.
It is natural that the effect on B-mode of GWs from kinks and that of GWs by global dynamics of strings are comparable.
The network dynamics can induce B-mode also through the vector perturbation it produces .
We expect that its contribution is also comparable to that of GWs from kinks, although strict comparison is difficult and requires numerical calculation.
The magnitude of the spectrum is proportional to $(G\mu)^2$,  
and hence when we take a different string tension
the shape is unchanged but the whole spectrum moves upward or downward.
If $G\mu \gtrsim \mathrm{(a~few)}\times 10^{-7}$, this spectrum can be observed by PLANCK, 
and if $G\mu \gtrsim  10^{-8}$, it can be detected by CMBpol.

The peak around $\ell \sim 5$ associated with the reionization has a characteristic shape.
However, it might be an artifact of the approximation that we put $\Psi$ out of the time integral in (\ref{aBlm}), assuming that $P(t)$ has a sharp peak around the reionization\footnote{
This approximation is much more valid around the recombination than the reionization, therefore the peak associated with the recombination does not have bump-like feature.
}
(see (\ref{Psiext})).
The actual peak might be smoother.
In fact, B-mode is induced over the finite time around the reionization, and those which are produced at different moments have their peak at different $\ell$.
It is expected that the total BB spectrum, which is the envelope of such peaks, has a smoother shape.
On the other hand, we expect that FIG. 1 shows the correct position and height of the peak  around $\ell \sim 5$.

  \section{conclusion}

In this paper, we have studied the effect of the stochastic GW background induced by kinks on the infinite cosmic strings on the BB power spectrum of CMB polarization.
Using the GW background obtained in Ref.~\cite{Kawasaki:2010yi}, 
we have estimated the resulting BB power spectrum.
We found that this effect is comparable to that of the vector and tensor modes induced by motion of the cosmic string network and may leave observable signatures in the spectrum.
If the cosmic string tension is large enough, 
the BB power spectrum by cosmic strings will be detected by future/on-going satellite experiments
such as PLANCK and CMBpol.
If it is discovered by the CMB experiments, then the direct detection of GWs 
from cosmic strings by pulsar timing arrays or 
space-laser interferometers may further confirm the existence of the cosmic string~\cite{Kawasaki:2010yi}.

\appendix 
\def\thesection{}
\section{}

In this Appendix, we derive Eq.~(\ref{CBB}) using the 
$\delta$-function-like property of the visibility function $P(t)$.
$\Psi(t,\qv,\lambda)$ varies more slowly than $P(t)$ at its peaks.
Therefore, the time integral in Eq.~(\ref{aBlm}) is approximated as
\be
	\int ^{t_0}_0 dt P(t) \Psi(t,\qv,\lambda) \chi_{\ell}(qr(t)) \simeq 
	\Psi(t_{\mathrm{rec}},q,\lambda) \int_{\mathrm{rec}} dt P(t)\chi_{\ell}(qr(t)) 
	+ \Psi(t_{\mathrm{rei}},q,\lambda) \int_{\mathrm{rei}} dt P(t) \chi_{\ell}(qr(t)), 
\tag{A1}
\label{Psiext}
\ee
where $\int_{\mathrm{rec}}(\int_{\mathrm{rei}})$ represents integration around the recombination(reionization), and $\chi_{\ell}(x)=(8x+x^2\partial/\partial x)(j_\ell(x)/x^2)$.
Then let us estimate $\Psi(t_{\mathrm{rec}},\qv,\lambda)$ and $\Psi(t_{\mathrm{rei}},\qv,\lambda)$ from Eq.~(\ref{Psieq}).
First, we consider $\Psi(t_{\mathrm{rec}},\qv,\lambda)$.
The factor $\exp(-\int^{t_{\mathrm{rec}}} _{t^{\prime}} dt^{\prime\prime} \omega_c(t^{\prime\prime}))$ has the property that it rapidly increases from $0$ when $t^{\prime}$ approaches $t_{\mathrm{rec}}$.
The combination $\omega_c(t^{\prime})\exp(-\int^{t_{\mathrm{rec}}} _{t^{\prime}} dt^{\prime\prime} \omega_c(t^{\prime\prime}))$ also has such a property.
while $\dot{\cD}$ and $\Psi$ vary more slowly than these functions.
Then, we obtain
\begin{align}
	\Psi(t_{\mathrm{rec}},\qv,\lambda) \simeq -3 & \dot{\cD}(t_{\mathrm{rec}},\qv,\lambda)  
	\int _{\mathrm{rec}}dt^{\prime}
	\exp\left(-\int^{\trec}_{t^{\prime}} dt^{\prime\prime}\omega_c(t^{\prime\prime})\right) 
	K\left(q\int^{\trec}_{t^{\prime}}\frac{dt^{\prime\prime}}{a(t^{\prime\prime})}\right) \nonumber \\
	& +\frac{3}{2}\Psi(\trec,\qv,\lambda)
	\int_{\mathrm{rec}}dt^{\prime}\omega_c(t^{\prime})
	\exp\left(-\int^{\trec}_{t^{\prime}}dt^{\prime\prime}\omega_c(t^{\prime\prime})\right)
	F\left(q\int^{\trec}_{t^{\prime}}\frac{dt^{\prime\prime}}{a(t^{\prime\prime})}\right).
	\tag{A2}
\end{align}
In contrast, $\exp(-\int^{t_{\mathrm{rei}}} _{t^{\prime}} dt^{\prime\prime} \omega_c(t^{\prime\prime}))$ is almost $1$ around $t^{\prime}=\trei$, since $\omega_c$ does not increase enough around the reionization epoch.
However, we can put $\dot{\cD}$ out of  the time integral because of the property of 
$K\left(q\int^{\trei}_{t^{\prime}}\frac{dt^{\prime\prime}}{a(t^{\prime\prime})}\right) $.
$K(x)$ decreases proportional to $x^{-3}$ for large $x$.
Besides, for $q>a(\trei)/\trei$, for which $\dot{\cD}(\trei,\qv,\lambda)$ has a nonzero value, $q\int^{\trei}_{t^{\prime}}\frac{dt^{\prime\prime}}{a(t^{\prime\prime})} \simeq 3\frac{q\trei}{a(\trei)}\left(1-\left(\frac{t^{\prime}}{\trei}\right)^{1/3}\right)$ grows rapidly when $t^{\prime}$ goes away from $\trei$.   
After all, $K\left(q\int^{\trei}_{t^{\prime}}\frac{dt^{\prime\prime}}{a(t^{\prime\prime})}\right)$ has a sharp peak at $t^{\prime}=\trei$.
Thus we can estimate $\Psi(t_{\mathrm{rei}},\qv,\lambda)$ as above,
\begin{align}
	\Psi(t_{\mathrm{rei}},\qv,\lambda) \simeq -3 & 
	\dot{\cD}(t_{\mathrm{rei}},\qv,\lambda)  
	\int _{\mathrm{rei}}dt^{\prime}
	\exp\left(-\int^{\trei}_{t^{\prime}} dt^{\prime\prime}\omega_c(t^{\prime\prime})\right) 
	K\left(q\int^{\trei}_{t^{\prime}}\frac{dt^{\prime\prime}}{a(t^{\prime\prime})}\right) \nonumber \\
	& +\frac{3}{2}\Psi(\trei,\qv,\lambda)\int_{\mathrm{rei}}dt^{\prime}\omega_c(t^{\prime})
	\exp\left(-\int^{\trei}_{t^{\prime}}dt^{\prime\prime}\omega_c(t^{\prime\prime})\right)
	F\left(q\int^{\trei}_{t^{\prime}}\frac{dt^{\prime\prime}}{a(t^{\prime\prime})}\right).
	\tag{A3}
\end{align}
Connecting the above estimations, we finally get
\be
	C_{BB,\ell}\simeq
	\pi^2T_0^2\int dqq^2\left[ A_\ell(q)\dot{\tilde{\cD}}^2(t_{\mathrm{rec}},q)
	+B_\ell(q)\dot{\tilde{\cD}}^2(t_{\mathrm{rei}},q)\right], 
	\tag{A4}
\ee
where 
\be
A_\ell(q)=\left( \frac{C(q)}{1-D(q)}\right)^2\times\left(\int_{\mathrm{rec}}dtP(t)\chi_{\ell}(qr(t))\right)^2, \  C(q)=-3\int_{\mathrm{rec}} dt^{\prime}
	e^{-\int^{\trec}_{t^{\prime}}dt^{\prime\prime}\omega_c(t^{\prime\prime})}
	K\left(q\int^{\trec}_{t^{\prime}}\frac{dt^{\prime\prime}}{a(t^{\prime\prime})}\right), \nonumber
	\ee
	\be
		\ D(q)=\frac{3}{2}\int_{\mathrm{rec}} dt^{\prime}
	\omega_c(t^{\prime})e^{-\int^{\trec}_{t^{\prime}}dt^{\prime\prime}\omega_c(t^{\prime\prime})}
	F\left(q\int^{\trec}_{t^{\prime}}\frac{dt^{\prime\prime}}{a(t^{\prime\prime})}\right), \nonumber
	\ee
and $B_\ell(q)$ is the function which we can get by substituting $\trec$ in $A_\ell(q)$ for $\trei$.

\begin{acknowledgments}

K.N. would like to thank the Japan Society for the Promotion of Science for financial support.
This work is supported by Grant-in-Aid for Scientific research from the Ministry of Education,
Science, Sports, and Culture (MEXT), Japan, No.14102004 (M.K.) 
and No. 21111006(M.K. and K.N.)
and also by World Premier International
Research Center Initiative (WPI Initiative), MEXT, Japan.

\end{acknowledgments}


 \end{document}